\documentclass[english, aps, prd, superscriptaddress,
	preprintnumbers,nofootinbib, twocolumn,
]{revtex4-1}

\usepackage[T1]{fontenc}
\usepackage[utf8]{inputenc}
\usepackage{verbatim}
\usepackage{empheq}
\usepackage{amssymb}
\usepackage{amsmath}
\usepackage{amsfonts}

\usepackage[colorlinks]{hyperref}
\hypersetup{citecolor=blue,linkcolor=blue,urlcolor=blue}

\usepackage{dcolumn}
\usepackage{graphicx}
\usepackage{etoolbox}
\usepackage{booktabs}
\usepackage{babel}

\newcommand{\LNP}{ {\Lambda} }
\newcommand{\ca}{\mathcal}
\newcommand{\eqspace}{\hphantom{{}={}}}
\newcommand{\cone}{{c_{HVV}^{(1)}}}
\newcommand{\ctwo}{{c_{HVV}^{(2)}}}
\newcommand{\ctilde}{{\tilde{c}^{\vphantom{(1)}}_{HVV}}}



\begin{document}
	
\global\long\def\order#1{\mathcal{O}\left(#1\right)}
\global\long\def\d{\mathrm{d}}
\global\long\def\P{P}
\global\long\def\amp{{\mathcal M}}
\preprint{OUTP-22-08P}
\preprint{TTP22-040}
\preprint{P3H-22-062}
\preprint{TIF-UNIMI-2022-9}

\def\BNL{ High Energy Theory Group, Physics Department, Brookhaven 
	National Laboratory, Upton, NY 11973, USA }
\def\KIT{ Institute for Theoretical Particle Physics, KIT, Karlsruhe, Germany }
\def\OX{ Rudolf Peierls Centre for Theoretical Physics, Clarendon Laboratory, 
	Parks Road, Oxford OX1 3PU, UK \\ and Wadham College, Oxford OX1 3PN, UK }
\def\MIL{ Tif Lab, Dipartimento di Fisica, Università di Milano and INFN, 
	Sezione di Milano, Via Celoria 16, I-20133 Milano, Italy}

\title{Anomalous Higgs boson couplings in weak boson fusion production at NNLO in QCD} 

\author{Konstantin~Asteriadis}
\email[Electronic address: ]{kasteriad@bnl.gov}
\affiliation{\BNL}

\author{Fabrizio~Caola}
\email[Electronic address: ]{fabrizio.caola@physics.ox.ac.uk}
\affiliation{\OX}

\author{Kirill~Melnikov}
\email[Electronic address: ]{kirill.melnikov@kit.edu}
\affiliation{\KIT}

\author{Raoul~R\"ontsch }
\email[Electronic address: ]{raoul.rontsch@unimi.it}
\affiliation{\MIL}

\begin{abstract}	
  The production of Higgs bosons in weak boson fusion has the second
  largest cross section among Higgs-production processes at the
  LHC. As such, this process plays an important role in detailed
  studies of Higgs interactions with vector bosons.  In this paper we
  extend the available description of Higgs boson production in weak
  boson fusion by considering anomalous $HVV$ interactions and NNLO
  QCD radiative corrections at the same time. We find that, while
  leading order QCD predictions are too uncertain to allow for
  detailed studies of the anomalous couplings, NLO QCD results are
  sufficiently precise, most of the time. The NNLO QCD corrections
  alter the NLO QCD predictions only marginally, but their
  availability enhances the credibility of conclusions based on NLO
  QCD computations.
\end{abstract}

\maketitle


\section{Introduction}
\label{sec:introduction}

 The discovery of the Higgs boson nearly ten years ago completed the
 Standard Model (SM) of particle physics, providing, for the first
 time, experimental support for the hypothesis that electroweak
 symmetry is broken by a scalar field.  By now quantum numbers of the
 discovered Higgs boson, as well as its couplings to gauge and matter
 fields, have been studied in great detail and it appears that the
 properties of the Higgs boson are closely aligned with the SM
 expectations.  For example, the Higgs couplings to massive
 electroweak vector bosons have been measured to within ${\cal O}(30)$
 percent of their Standard Model values~(see
 e.g. Refs~\cite{ATLAS:2019nkf,CMS:2021nnc}).

 Verifying the Higgs boson couplings to a better, perhaps a few
 percent, precision is the goal of the LHC Run III and, especially, of
 the high-luminosity LHC.  Reaching this goal will not be easy as it
 will require combining precise measurements with very detailed
 theoretical predictions.  In addition, assuming that deviations in
 Higgs couplings are found, understanding their origin and what they
 imply becomes important. A convenient way to investigate this in a
 relatively model-independent way is to use an Effective Field Theory
 (EFT) framework and parameterize deviations from the Standard Model
 by higher-dimensional operators that depend on the Standard Model
 fields. These operators modify interactions between Standard Model
 particles and affect production cross sections and kinematic
 distributions that are observed at the LHC. An EFT that extends the
 SM is known as
 SMEFT~\cite{Buchmuller:1985jz,Grzadkowski:2010es,Brivio:2017vri}.

 Assuming that SMEFT provides a faithful description of beyond the
 Standard Model (BSM) physics, it becomes important to extract Wilson
 coefficients of higher-dimensional operators from the experimental
 data. Similar to measurements of Standard Model parameters, such an
 extraction of Wilson coefficients is affected by radiative
 corrections, of which QCD corrections are especially important.  In
 this paper we investigate the interplay between anomalous couplings
 and radiative corrections for Higgs production in weak boson fusion
 (WBF).  Such an interplay can be quite subtle for this
 process. Indeed, corrections to inclusive Higgs production in WBF are
 known to be small, but their impact on kinematic distributions can be
 larger.  Anomalous couplings also distort kinematic distributions
 and, in fact, it is known that shapes of various observables often
 provide the best means to distinguish between different anomalous
 couplings (see
 e.g. Refs~\cite{Plehn:2001nj,Figy:2004pt,Hankele:2006ma}).  Hence,
 for a reliable EFT analysis it becomes important to understand to
 what extent the effects of anomalous couplings and QCD corrections
 can be disentangled.

 The theoretical description of Higgs production in weak boson fusion
 is very advanced. At the inclusive level, N$^3$LO QCD corrections are
 known in the so-called factorized
 approximation~\cite{Dreyer:2016oyx}.  At the differential level,
 factorized NNLO QCD corrections~
 \cite{Cacciari:2015jma,Cruz-Martinez:2018rod,Asteriadis:2021gpd} as
 well as NLO EW
 corrections~\cite{Ciccolini:2007jr,Ciccolini:2007ec,Figy:2010ct} are
 available. The dominant non-factorizable corrections have been
 studied in
 Refs~\cite{Bolzoni:2010xr,Bolzoni:2011cu,Liu:2019tuy,Dreyer:2020urf}.
 Theoretical predictions that include the anomalous couplings together
 with the NLO corrections to Higgs boson production in weak boson
 fusion can be obtained using such programs as
 HAWK~\cite{Denner:2014cla}, VBFNLO
 \cite{Hankele:2006ma,Baglio:2014uba} and
 Madgraph5~\cite{Alwall:2011uj}.

 In this paper we extend these results by computing NNLO QCD
 corrections to weak boson fusion in the factorized approximation in
 the presence of anomalous $HVV$ couplings.  Although the $HVV$ vertex
 is not affected by QCD effects, such a computation is non-trivial
 since obtaining fully-differential predictions for the WBF process
 with NNLO QCD accuracy
 \cite{Cacciari:2015jma,Cruz-Martinez:2018rod,Asteriadis:2021gpd} is
 quite demanding.  The reason behind this is the relative smallness of
 radiative corrections and a very large phase space (typical for a $2
 \to 5$ process) whose efficient generation is challenging.

 In this paper, we employ the calculation of NNLO QCD corrections
 reported in Ref.~\cite{Asteriadis:2021gpd}. It is performed using the
 nested soft-collinear subtraction scheme~\cite{Caola:2017dug} which
 was adapted to WBF kinematics in Ref.~\cite{Asteriadis:2019dte}. It
 provides a sufficiently efficient implementation and phase-space
 sampling that allowed us to incorporate decays of the Higgs
 boson~\cite{Asteriadis:2021gpd} into the calculation.  This
 efficiency is quite important for studying the anomalous couplings,
 especially for performing scans in the parameter space.

 The rest of the paper is organized as follows. In
 Section~\ref{sec:anom_feynman_rule} we describe the EFT framework
 which we employed in the calculations reported in this paper, and
 discuss the parametrization of the anomalous couplings adopted for
 this analysis.  In Section~\ref{sec:results} we present
 phenomenological results for the 13 TeV LHC. After describing our
 setup, we show predictions for cross sections and scans in the space
 of the anomalous couplings.  We then focus on two scenarios where
 anomalous interactions are present but fiducial cross sections are
 indistinguishable from their Standard Model values.  We then consider
 kinematic distributions which are sensitive to the anomalous
 couplings of the Higgs boson, and discuss the impact of higher order
 QCD corrections to them.  We conclude in Section~\ref{sec:concl}.
    
\section{Anomalous HVV interactions}
\label{sec:anom_feynman_rule}
For our analysis, we focus on anomalous $HVV$ interactions, where $V$
is a massive vector boson.\footnote{For definiteness, we do not
  consider photon-mediated WBF in our study. We stress however that
  its inclusion in our framework is straightforward.}  Lorentz
invariance and Bose symmetry imply that the most generic $HVV$ vertex
must have the form\footnote{We note that, in principle, tensor
  couplings of the form $p_1^\mu p_2^\nu$ are allowed in
  Eq.~(\ref{eq:feynman:hvvw}). However, neglecting contributions of
  heavy quarks in $q \to q V$ and similar transitions, the vector
  bosons couple to conserved currents in weak boson fusion, so that
  such couplings do not contribute to the final result.}
\begin{align}
	\label{eq:feynman:hvvw}
	\begin{split}
		\hspace{-5pt}\vcenter{\hbox{\includegraphics[scale = 0.39]{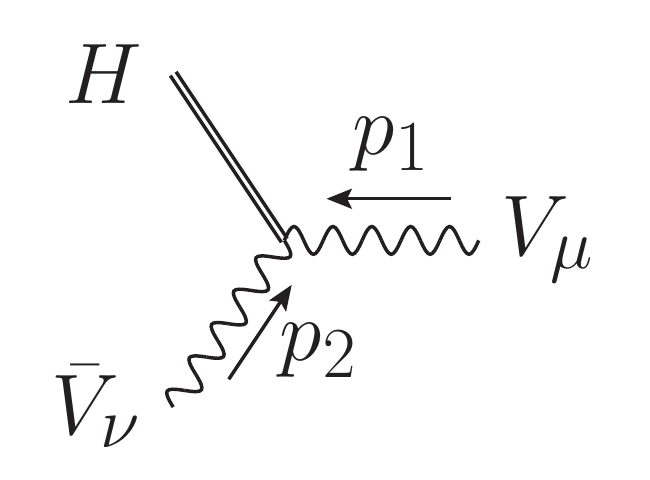}}} \hspace{-2pt} &= \hspace{5pt}
		i \Big[ g^{\mu\nu} A(p_1^2,p_2^2,p_1\cdot p_2)  \\[-16pt]
		&+ p_1^\nu p_2^\mu\; B(p_1^2,p_2^2,p_1\cdot p_2) \\
		&+ i \epsilon^{\mu\nu\rho\sigma}  p_{1,\rho}p_{2,\sigma}\;
		C(p_1^2,p_2^2,p_1\cdot p_2) 
		\Big] \, ,
	\end{split}
\end{align}
where $A$,$B$ and $C$ are arbitrary functions which, in the context of
an effective theory, are Taylor-expandable in $p_i^2$ and $p_i\cdot
p_j$ ($i,j=1,2$).  It turns out that the dependence on these kinematic
invariants can be further restricted. This point has been studied
quite recently through an all-order EFT expansion in the context of
the Standard Model Effective Theory~\cite{Helset:2020yio}.  In case of
the dominant dimension-six operators, it has been
shown~\cite{Helset:2020yio} that the functions $A$,$B$ and $C$ can be
written as follows
\begin{align}
	\begin{split}
		A &= g_{HVV}^{(SM)} + g_{HVV}^{(3)} + g^{(1)}_{HVV} \frac{p_1 \cdot p_2}{\Lambda^2} \, , \\
		B &= -\frac{g^{(1)}_{HVV}}{\Lambda^2} \, , \qquad
		C = \frac{i\tilde g_{HVV}}{\Lambda^2} \, ,
	\end{split}
	\label{eq:ABC}
\end{align}
where $\Lambda$ is the EFT scale and $g_{HVV}^{(i)}$ are
$p_{1,2}$-independent constants.  In Eq.~(\ref{eq:ABC}),
$g_{HVV}^{(SM)}$ is the Standard Model coupling, i.e. $g_{HWW}^{(SM)}
= g\, m_W$ and $g_{HZZ}^{(SM)} = g\, m_Z / \cos \theta_W$, where $g$
is the weak coupling and $\theta_W$ is the weak mixing
angle.\footnote{We note that the $g_{HVV}^{(i)}$ couplings introduced
  in Eq.~\eqref{eq:ABC} are defined exactly as in
  Ref.~\cite{Mimasu:2015nqa}. In this reference, an additional
  coefficient $g_{HVV}^{(2)}$ is also present. However, this Wilson
  coefficient is not generated at dimension six~\cite{Helset:2020yio}.
  For this reason we do not consider it in our analysis, although it
  is straightforward to include it if needed.}

We find it convenient to parametrize the anomalous couplings as follows 
\begin{align}
\begin{split} 
  & \frac{g^{(1)}_{HVV}}{g^{(SM)}_{HVV}}  = 2  c^{(1)}_{HVV} \, , \qquad \frac{\tilde g_{HVV}}{g_{HVV}^{(SM)}} = 6 \pi \tilde c_{HVV} \, ,
  \\
  & \frac{g^{(3)}_{HVV} }{g_{HVV}^{(SM)}} + \frac{m_H^2}{\Lambda^2} \frac{g^{(1)}_{HVV}}{g^{(SM)}_{HVV}} = \frac{m_H^2}{\Lambda^2} c^{(2)}_{HVV} \, .
\end{split} 
\end{align}
The new constants $c_{HVV}^{(i)}$ are dimensionless; in terms of these
constants, the coupling of the Higgs boson and the electroweak vector
bosons reads, c.f.  Eq.~\eqref{eq:feynman:hvvw},
\begin{align}
	\label{eq:feynman:hvvw2}
	\begin{split}
		\hspace{-5pt}\vcenter{\hbox{\includegraphics[scale=0.39]{figs/hvv_vertex.pdf}}} \hspace{-2pt} &= \hspace{5pt}  i
		g_{HVV}^{(SM)} \bigg[ g^{\mu\nu} \bigg( 1 + \frac{
			m_H^2}{\LNP^2} \ c_{HVV}^{(2)} \\[-12pt]
		&\eqspace+ \frac{p_1^2 + p_2^2}{\LNP^2} \ c_{HVV}^{(1)} \bigg) + \frac{2 p_1^\nu
			p_2^\mu}{\LNP^2} c_{HVV}^{(1)} \nonumber
		\end{split}\\
		&- \tilde{c}_{HVV} (6\pi)
		\epsilon^{\mu\nu\rho\sigma}
		\frac{p_{1,\rho}p_{2,\sigma}}{\LNP^2} \bigg] \, .
\end{align}
The factor $(6 \pi)$ introduced in the $CP$-odd term is chosen for
convenience; indeed, as we will see later, with this normalization and
for $\Lambda = 1~{\rm TeV}$, $\tilde c_{HVV} \sim 1$ induces ${\ca
  O}(1 \%)$ corrections to the fiducial cross section.

In general, the $HZZ$ and $HWW$ anomalous couplings do not have to be
the same. However, for simplicity, in the analysis below we will
assume that, by factorizing the SM couplings, we already account for
main differences between them. Hence, in what follows, we restrict
ourselves to the case where $c_{HZZ}^{(1,2)} =
c_{HWW}^{(1,2)}=c_{HVV}^{(1,2)}$ and $\tilde c_{HZZ} = \tilde c_{HWW}
= \tilde c_{HVV}$.

\section{Results at the 13 TeV LHC}
\label{sec:results}

\subsection*{Setup}
For the phenomenological results reported in this paper, we use the
same parameters and kinematic selection criteria as in
Ref.~\cite{Asteriadis:2019dte}. We collect them here for
completeness. We consider proton-proton collisions with the
center-of-mass energy 13 TeV and treat the Higgs boson as stable.  We
use the Higgs mass $m_H =125~{\rm GeV}$, the vector boson masses $M_W
= 80.398~{\rm GeV}$ and $M_Z = 91.1876~{\rm GeV}$, and their widths
$\Gamma_W = 2.105~{\rm GeV}$ and $\Gamma_Z = 2.4952~{\rm GeV}$. We use
the Fermi constant $G_F = 1.16639 \times 10^{-5} {\rm GeV}^{-2}$ to
derive the weak couplings and we set the CKM matrix to an identity
matrix.  We employ the \texttt{NNPDF31-nnlo-as-118} parton
distribution functions~\cite{NNPDF:2017mvq} and use them for all
calculations reported in this paper irrespective of the nominal
perturbative order.  We also use $\alpha_s(M_Z) = 0.118$.  The
evolution of the parton distribution functions and the strong coupling
constant is obtained from LHAPDF~\cite{Buckley:2014ana}.  Finally, we
employ dynamical renormalization and factorization scales $ \mu_R =
\mu_F = \mu$ using a central value~\cite{Cacciari:2015jma}
\begin{align}
  \mu_0 = \sqrt{ \frac{m_H}{2} \sqrt{
    \frac{m_H^2}{4} + p_{\perp,H}^2} } \, .
\label{eq:dynsc}
\end{align}
To define the weak boson fusion fiducial volume we reconstruct jets
using the inclusive anti-$k_\perp$ algorithm \cite{Cacciari:2008gp}
with $R = 0.4$.  We require events to contain at least two jets with
transverse momenta $p_{\perp,j} > 25~{\rm GeV}$ and rapidities $|y_j|
< 4.5$. Also, the two leading-$p_\perp$ jets should be separated by a
large rapidity interval $|y_{j_1} - y_{j_2}| > 4.5$ and their
invariant mass should be larger than $600~{\rm GeV}$. In addition, the
two leading jets should be in different hemispheres in the laboratory
frame; to enforce this, we require that the product of their
rapidities in the laboratory frame is negative, $y_{j_1} y_{j_2} < 0$.
Finally, for definiteness we set $\Lambda = 1~{\rm TeV}$ in all
computations that we report below.

\subsection*{Fiducial cross sections}
\label{sec:fid}

\begin{table}[t]
  \begin{center}
    \begin{tabular}{lccr}
      \toprule
      $\sigma_{\rm fid}$ (fb) \hspace{-10pt} & LO & NLO  & NNLO  \\
      \midrule 
      $X_1$ & $971_{+69}^{-61}$ & $890_{-18}^{+8}$ & $859_{-10}^{+8} $ \\[3pt]
      $X_2$ & \hspace{-10pt} $0.413_{+0.039}^{-0.033}$ \hspace{-10pt} & $0.398_{-0.005}^{-0.001}$ & $0.383_{-0.005}^{+0.004}$ \\[3pt]
      $X_3$ & \hspace{-10pt}  $19.57_{+2.22}^{-1.84}$ \hspace{-10pt}  & $19.64_{-0.07}^{-0.25}$ & $ 19.25_{-0.18}^{+0.08} $ \\[3pt]
      $X_4$ & \hspace{-10pt}  $26.43_{+1.80}^{-1.61}$ \hspace{-10pt}  & $23.45_{-0.66}^{+0.35}$ & $22.53_{-0.42}^{+0.39}$ \\  
      \bottomrule
    \end{tabular}
    \caption{Fiducial cross sections, in fb, at various orders of
      perturbative QCD using $\mu = \mu_0$. The sub- and super-scripts
      indicate the results computed with $\mu = \mu_0/2$ and $\mu =
      2\mu_0$, respectively. Numerical uncertainties are much smaller
      than scale uncertainties and are not shown.  See text for
      further details.  \label{tab:n1}}
  \end{center}
\end{table}

We now present our results for the cross sections for Higgs production
in weak boson fusion in the fiducial region described above, through
NNLO QCD.  Since three independent couplings appear in the $HVV$
vertex, cf. Eq.~\eqref{eq:feynman:hvvw2}, the cross section naturally
separates into six terms
\begin{align}
  \begin{split}
     \sigma_{\rm fid} &= \left(1+\frac{m_H^2}{\LNP^2}\ctwo\right)^2 X_1 +
    \left(\cone\right)^2 X_2 \\ & +
    \left(\ctilde\right)^2 X_3 + \left(1+ \frac{m_H^2}{\LNP^2}
    \ctwo \right) \cone X_4 \\ &+ \left(1+
    \frac{m_H^2}{\LNP^2} \ctwo\right) \ctilde X_5 + \tilde
    c_{HVV}^{(1)} \ctilde X_6 \, .
  \end{split}
  \label{eq2}
\end{align}
We note that $X_1$ is the Standard Model cross section.  Two of the
other terms, $X_{5,6}$ describe the interference of $CP$-even and
$CP$-odd couplings and hence vanish for the azimuthally-symmetric cuts
that we employ for computing fiducial cross sections.\footnote{Note,
  however, that this is \emph{not} the case for azimuthally-sensitive
  observables. It is therefore important not to discard these
  contributions altogether, as they can play an important role in
  kinematic distributions.}  The results for the non-vanishing
coefficients at leading, next-to-leading and next-to-next-to-leading
orders are shown in Table~\ref{tab:n1}.

We note that the radiative corrections to the various cross sections
$X_{1,2,3,4}$ are similar but not identical. In general, when moving
from LO to NLO we also observe a significant reduction in the
dependence of the cross sections $X_{1,2,3,4}$ on the renormalization
and factorization scales, which we estimate by changing the central
scale $\mu_0$ in Eq.~(\ref{eq:dynsc}) by a factor two in either
direction.  Instead, there is no substantial scale-uncertainty
reduction when moving from NLO to NNLO, and the NLO scale variation
bands do not contain the NNLO result. This feature is well-known for
the SM case, where it is also known that the NNLO and N$^3$LO scale
variation bands of the inclusive cross section do
overlap~\cite{Dreyer:2016oyx}. Because of this, it is understood in
the context of SM studies that drawing conclusions based only on NLO
QCD predictions and their scale uncertainty is delicate, while NNLO
analyses should be more robust. Table~\ref{tab:n1} suggests that this
also holds in the presence of anomalous $HVV$ interactions.  These
features will play an important role in the discussion below.

\begin{figure}
  \centering
  \hspace{-25pt}
  \includegraphics[width=200pt]{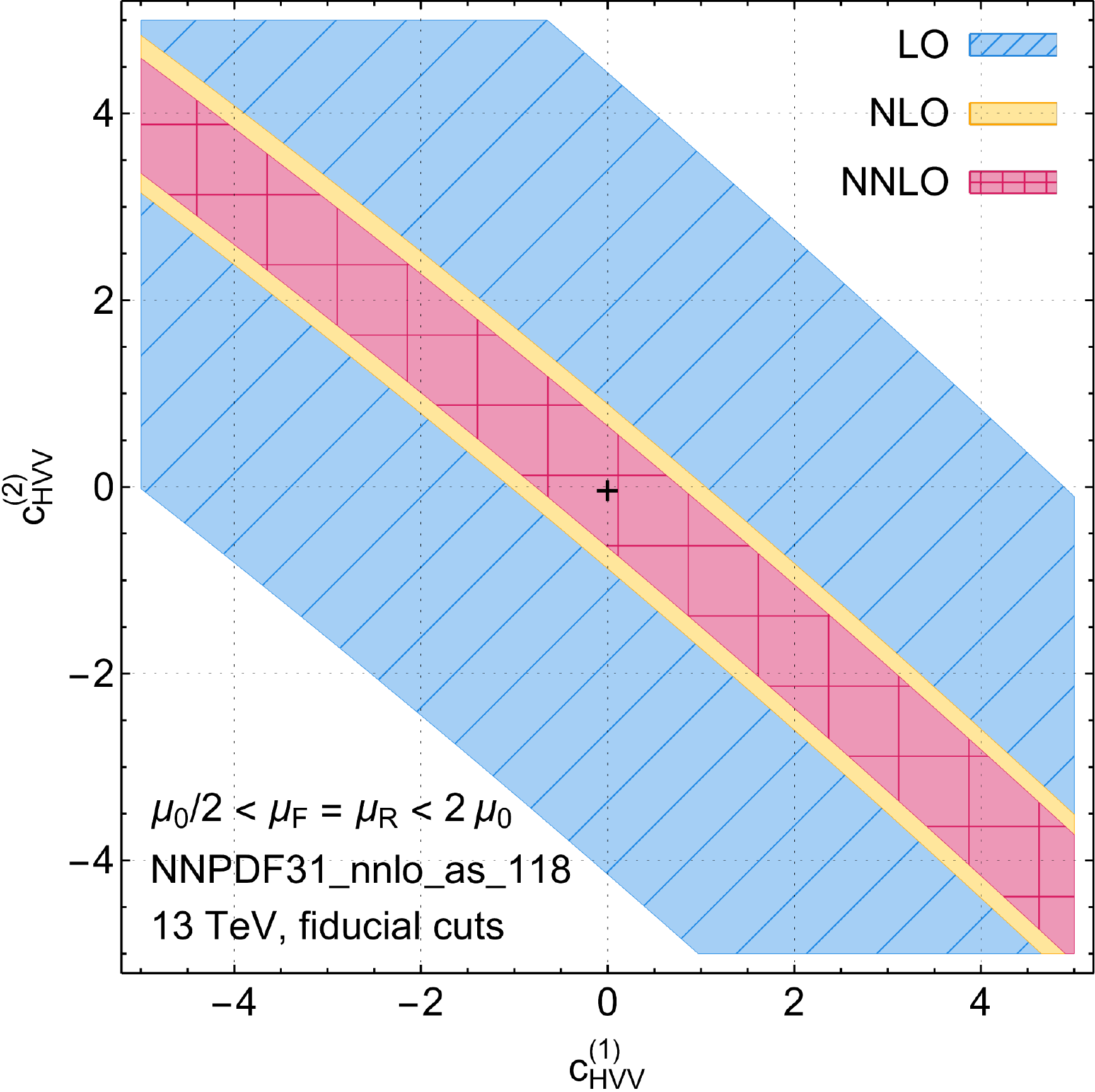}
  \caption{Allowed combinations of anomalous couplings $c^{(1)}_{HVV}$
    and $c^{(2)}_{HVV}$ where the residual scale uncertainty band
    overlaps with SM prediction at the 13 TeV LHC. The color coding
    describes LO (hashed blue), NLO (yellow) and NNLO (squared red)
    calculations. The SM result is shown as black cross. See text for
    details.}
  \label{fig:ellipses:cp_even}
\end{figure}

\begin{figure}[t]
	\centering
	\hspace{-25pt}
	\includegraphics[width=200pt]{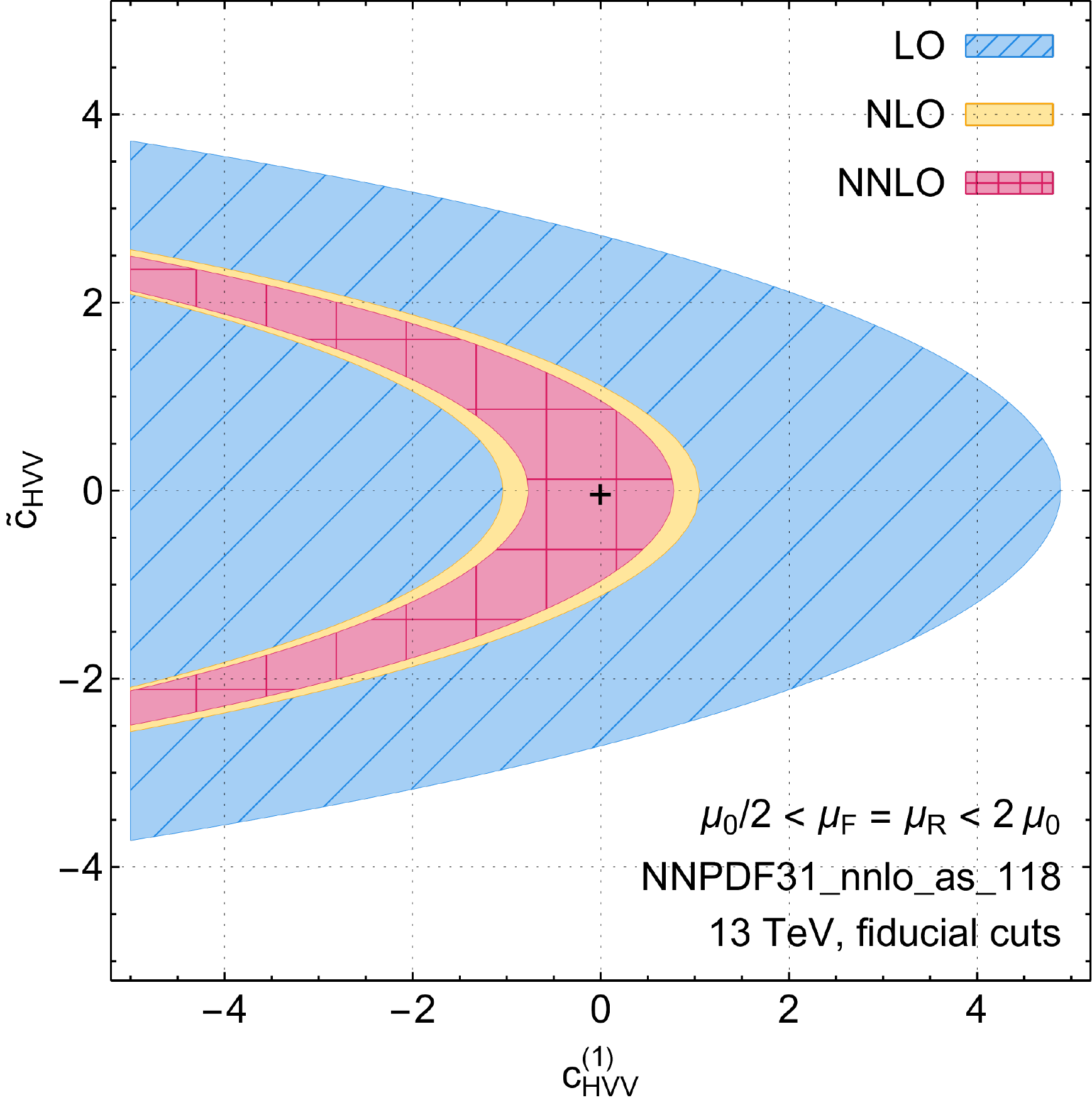}\\[10pt]
	\hspace{-25pt}
	\includegraphics[width=200pt]{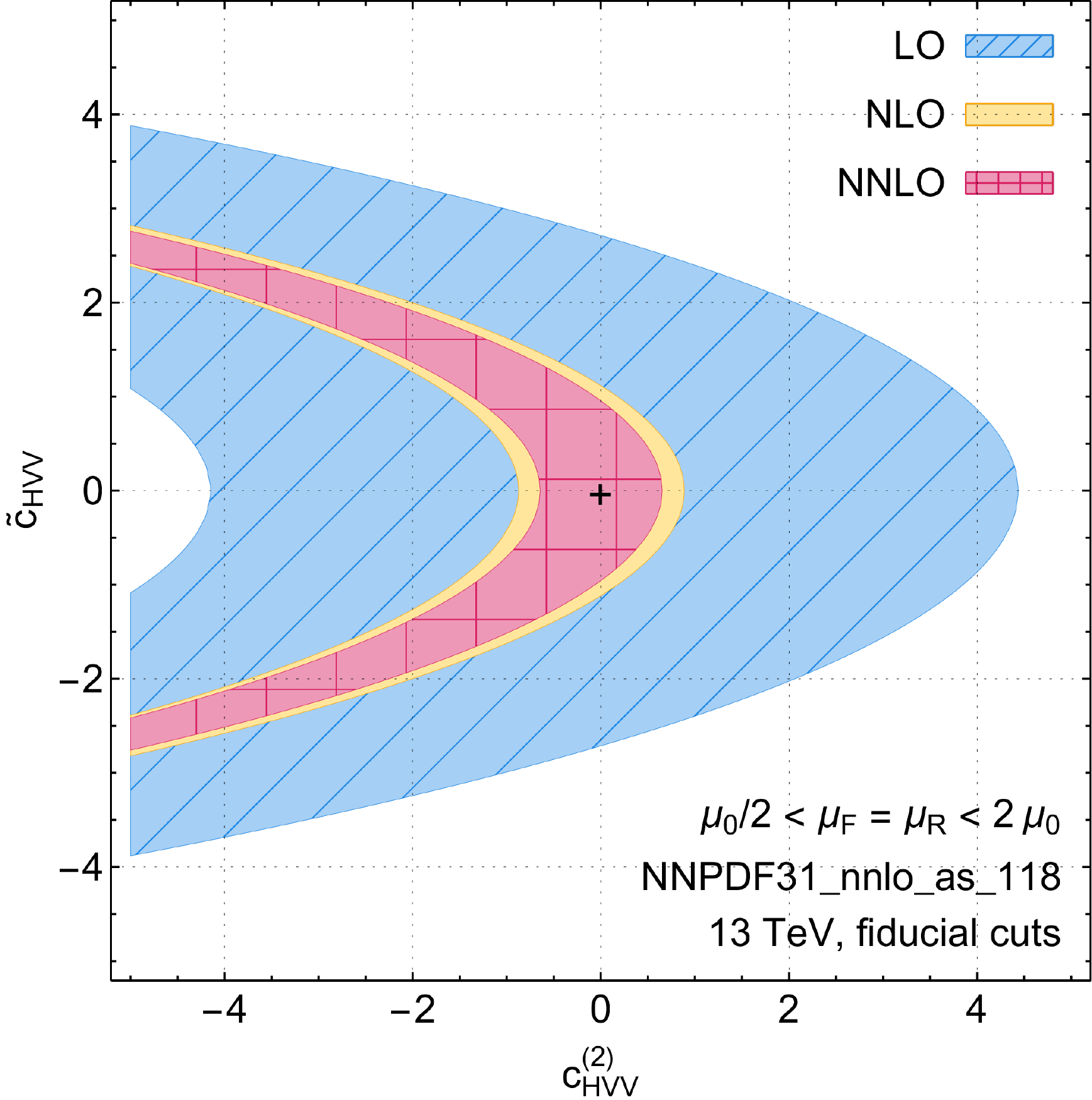}
	\caption{Same as Fig.~\ref{fig:ellipses:cp_even} but for
          combinations involving the CP-odd anomalous coupling
          $\tilde{c}_{HVV}$. See text for details.}
	\label{fig:ellipses:cp_odd}
\end{figure}

To illustrate the importance of higher order corrections, we now study
the dependence of fiducial cross sections on the anomalous coupling
constants at different orders of QCD perturbation theory. To this end,
we take the Standard Model cross section as a reference point. We then
consider a hypothetical BSM scenario with nonzero anomalous
couplings, and declare it to be compatible with the Standard Model if
the scale variation bands of the SM and BSM predictions overlap at a
given order in QCD.

We show the results of this analysis in
Figs~\ref{fig:ellipses:cp_even},\ref{fig:ellipses:cp_odd} where
different pairs of the anomalous couplings are chosen for
two-dimensional projections. In those figures, the colored regions
represent the allowed values for anomalous couplings according to the
criterion just explained. We note that for this analysis we take
Eq.~\eqref{eq2} exactly as written, i.e.  we include the contribution
of dimension-six operators squared. One may worry about the
consistency of this approach since we are neglecting the contribution
of dimension-eight operators when defining the anomalous
couplings. While this worry is clearly justified, the goal of our
analysis is to study the possible interplay between higher order QCD
corrections and EFT effects, and for this dealing with a
representative set of anomalous couplings is sufficient.  We stress
however that the results of Table~\ref{tab:n1} make it straightforward
for anyone to repeat the analysis including only the interference of
dimension-six and SM operators. Similarly, it is immediate to change
the definition of compatibility between SM and BSM predictions.

Having clarified our procedure, we now comment on the results,
c.f. Figs~\ref{fig:ellipses:cp_even},\ref{fig:ellipses:cp_odd}.  We
note that since the cross sections $X_{2,3,4}$ describing the
contributions of the anomalous couplings change the Standard Model
cross section by about $2-3$ percent and the scale uncertainty of the
leading order Standard Model cross section is about $10$~percent, only
relatively large anomalous couplings can be excluded at leading order.
The situation changes dramatically at NLO since theoretical
uncertainties are reduced to about $1-2$ percent. There are further
improvements at NNLO but they are minor; such improvements would imply
changes in the anomalous couplings at the level of $\delta
c_{HVV}^{(1,2)} \sim 0.1$ which, depending on the value of $c_{HVV}$
can be about ten percent. However, as we noted above, the NNLO shifts
in the cross sections shown in Table~\ref{tab:n1} are often outside
the NLO scale uncertainty bands. We then take the consistency of the
NLO and NNLO results as an indicator of the reliability of the NLO
analysis.

It is clear that the above study can be made more complex by, for
example, considering three couplings at the same time, or by comparing
results that include the contribution of dimension-six operators
squared and results that do not, or by defining a more refined
estimate of the compatibility of SM and BSM results.  We do not pursue
these investigations in this paper but we point out that once the
fiducial cross sections $X_{1,2,3,4}$ are known, defining selection
criteria and scanning the parameter space become straightforward.
Obviously, if the fiducial cuts are to be changed, then all the cross
sections $X_{1,2,3,4}$ have to be recalculated.  We note in this
respect that it takes about $25\,000$ CPU hours to compute each $X_i$
with good precision. Hence, if ${\cal O}(1000)$ CPU's are available,
about a few days of running is required.

\begin{figure*}[ht!]
  \centering
  \includegraphics[width=168pt]{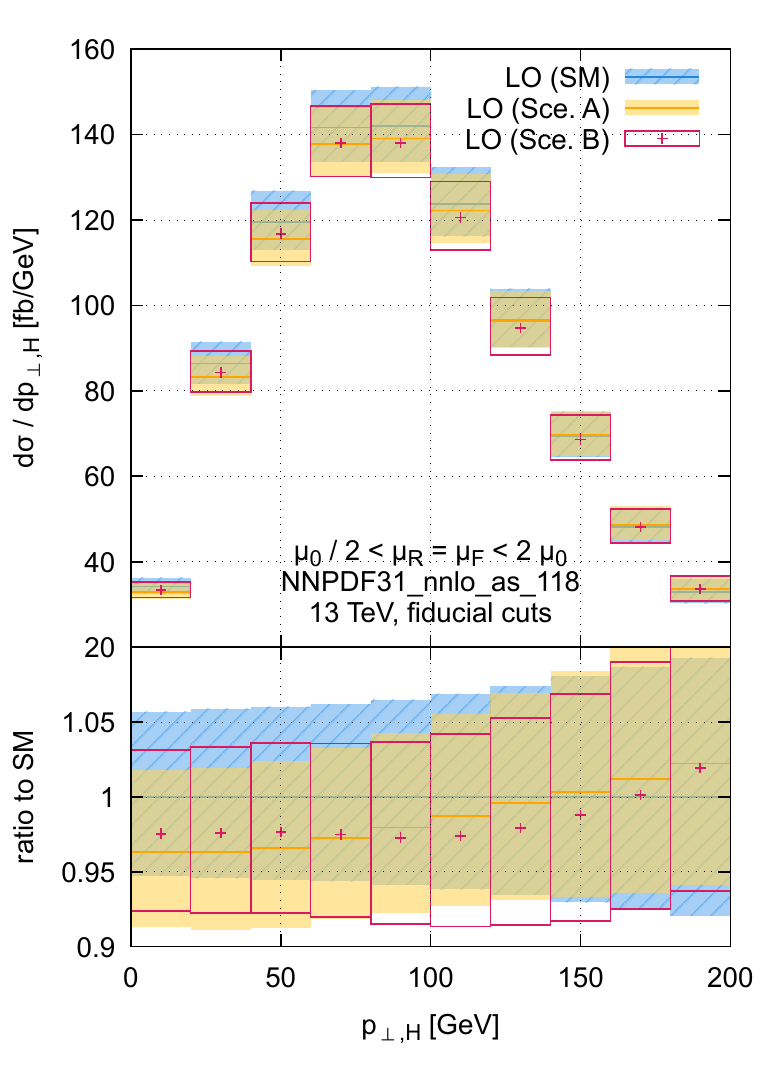}
  \includegraphics[width=168pt]{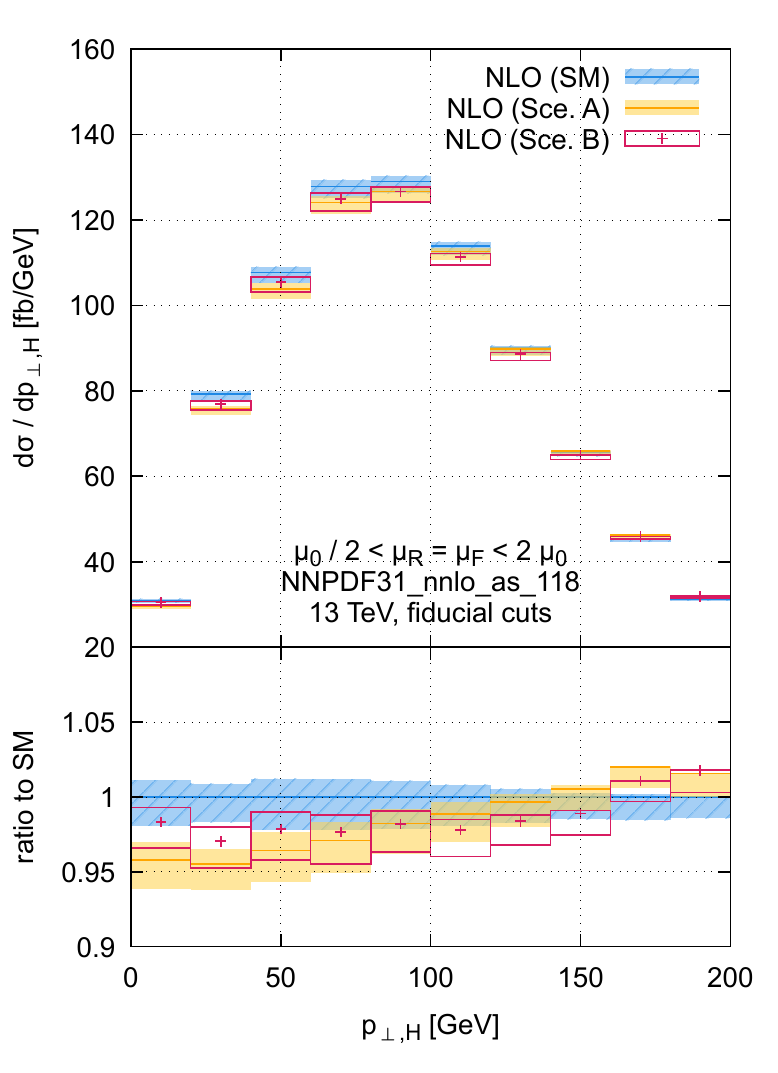}
  \includegraphics[width=168pt]{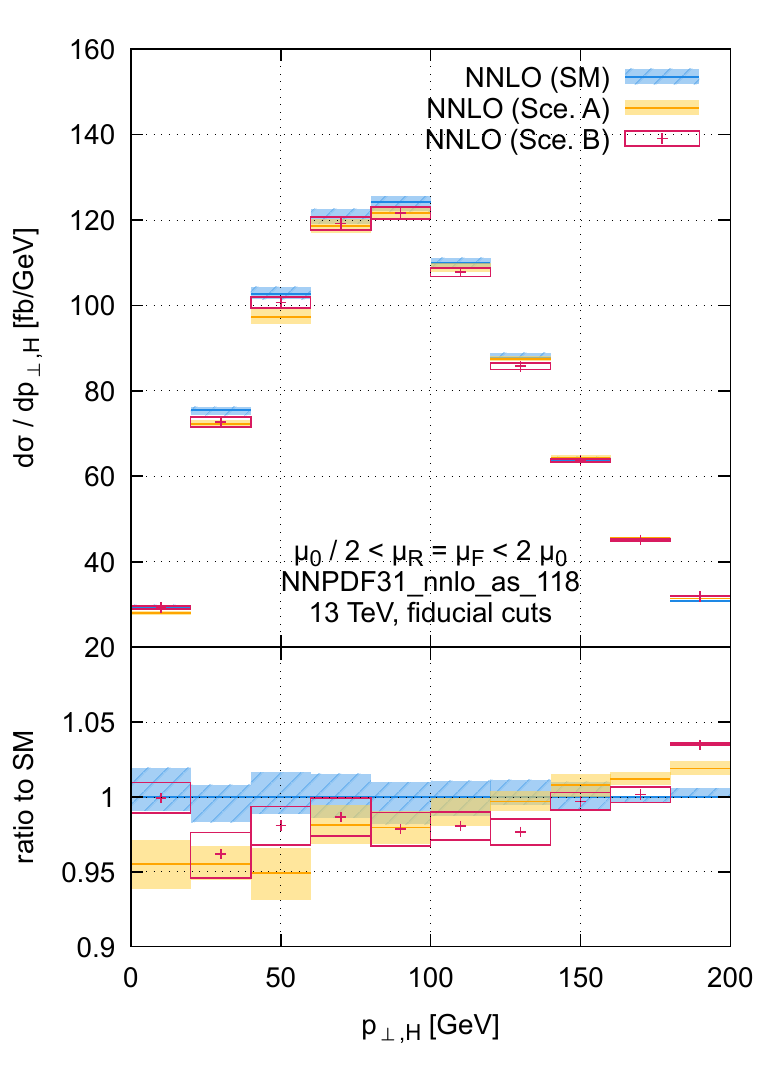}
  \caption{Transverse momentum distributions of the Higgs boson for LO
    (left), NLO (middle) and NNLO (right) QCD predictions. For each
    plot, the upper pane displays the SM (hashed blue), scenario A
    (solid yellow) and scenario B (red boxes) results as defined in
    the main text.  The lower pane shows the ratio of these results to
    the SM result at central scale. The lines indicate the central
    renormalization and factorization scale choice, and the bands
    indicate the envelope of the results at different scales. See text
    for details.}
  \label{fig:histo:scea:pth}
\end{figure*}

\subsection*{Kinematic distributions}

In the previous section, we discussed fiducial cross sections. This,
of course, does not exhaust all opportunities for the analysis as
kinematic distributions can also help to distinguish between the
different couplings.  To illustrate this point and to show how
higher-order QCD corrections impact such an analysis, we consider two
sets of anomalous couplings which lead to nearly identical cross
sections.  We choose the following scenarios
\begin{alignat*}{4}
\textrm{Sce. A:}& \ \ &c_{HVV}^{(1)} &= +1.5 \, , &\ \ c_{HVV}^{(2)} &= -1.9 \, , &\ \ \tilde c_{HVV} &= +0.6 \, ; \\
\textrm{Sce. B:}& \ \ &c_{HVV}^{(1)} &= -1.8 \, , &\ \ c_{HVV}^{(2)} &= -0.1 \, , &\ \ \tilde c_{HVV} &= -1.5 \, .
\end{alignat*}

The results for fiducial cross sections at various orders of QCD
perturbation theory are shown in Table~\ref{tab:sce:results}. It is
clear that, at leading order, the two scenarios cannot be
distinguished from each other and from the Standard Model.  Even with
the significant reduction of scale variation uncertainties at
next-to-leading and next-to-next-to-leading orders, the fiducial cross
sections remain compatible.

\begin{table}
\begin{center}
	\begin{tabular}{lccc}
	  \toprule
	  $\sigma_\textrm{fid}$ (fb) & SM & Sce.~A & Sce.~B \\
	  \midrule
	  LO & $971^{-61}_{+69}$ & $960^{-61}_{+68}$ & $965^{-63}_{+71}$ \\[3pt]
	  NLO & $890^{+8}_{-18}$ & $882^{+7}_{-17}$ & $890^{+6}_{-17}$   \\[3pt]
	  NNLO & $859^{+8}_{-10}$ & $851^{+9}_{-8}$ & $860^{+8}_{-8}$   \\
	  \bottomrule
	\end{tabular}
\end{center}
\caption{Central values of fiducial cross sections for the Standard
  Model and Scenarios A and B using $\mu = \mu_0$. The sub- and
  super-scripts indicate the results computed with $\mu = \mu_0/2$ and
  $\mu = 2\mu_0$, respectively. Numerical uncertainties are much
  smaller than scale uncertainties and are not shown.}
\label{tab:sce:results}
\end{table}

To understand if the two scenarios and the Standard Model can still be
distinguished from each other, we consider the kinematic
distributions. For most observables, however, there is very little
difference between the various scenarios even at higher orders.  To
give an example, in Fig.~\ref{fig:histo:scea:pth} we consider the
Higgs transverse momentum distribution.  We report predictions at LO
(left), NLO (middle) and NNLO (right), for the scenarios A and B and
the SM. It is clear from this figure that one cannot disentangle the
different models using LO predictions. At NLO and NNLO one starts
seeing hints of slightly different shapes, but this is a mild effect
and the residual scale variation uncertainty is still too large to
allow any definite conclusion to be drawn. We note that the
predictions for the three models start deviating significantly at
large transverse momenta $p_{\perp,H}\gtrsim 250$~GeV. This is
expected as the EFT effects are enhanced at higher energies. However,
in this region the cross section is already quite small, having
dropped off by about one order of magnitude from its peak value.

\begin{figure*}
	\centering
	\includegraphics[width=168pt]{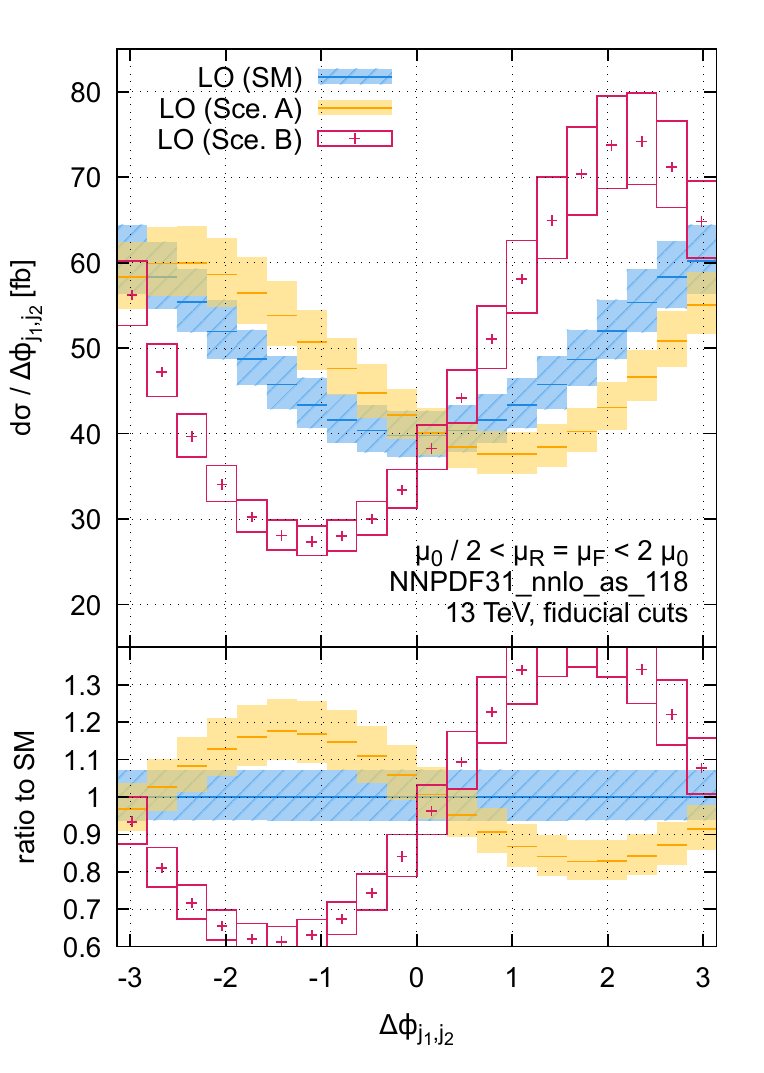}
	\includegraphics[width=168pt]{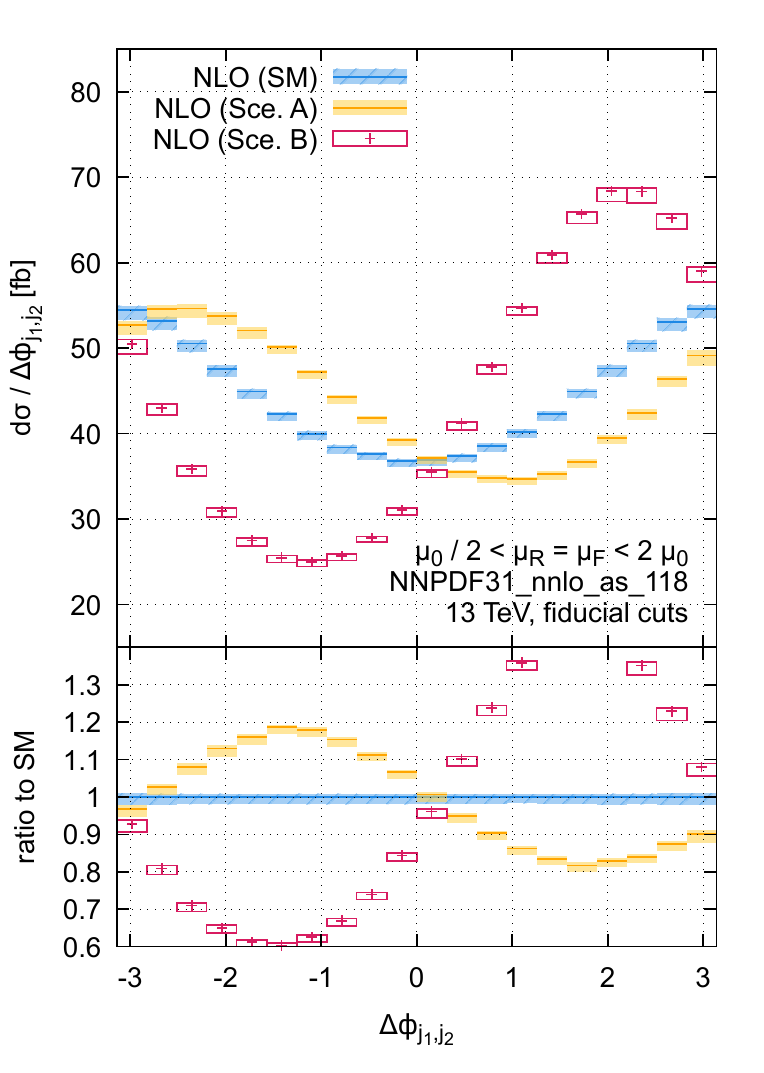}
	\includegraphics[width=168pt]{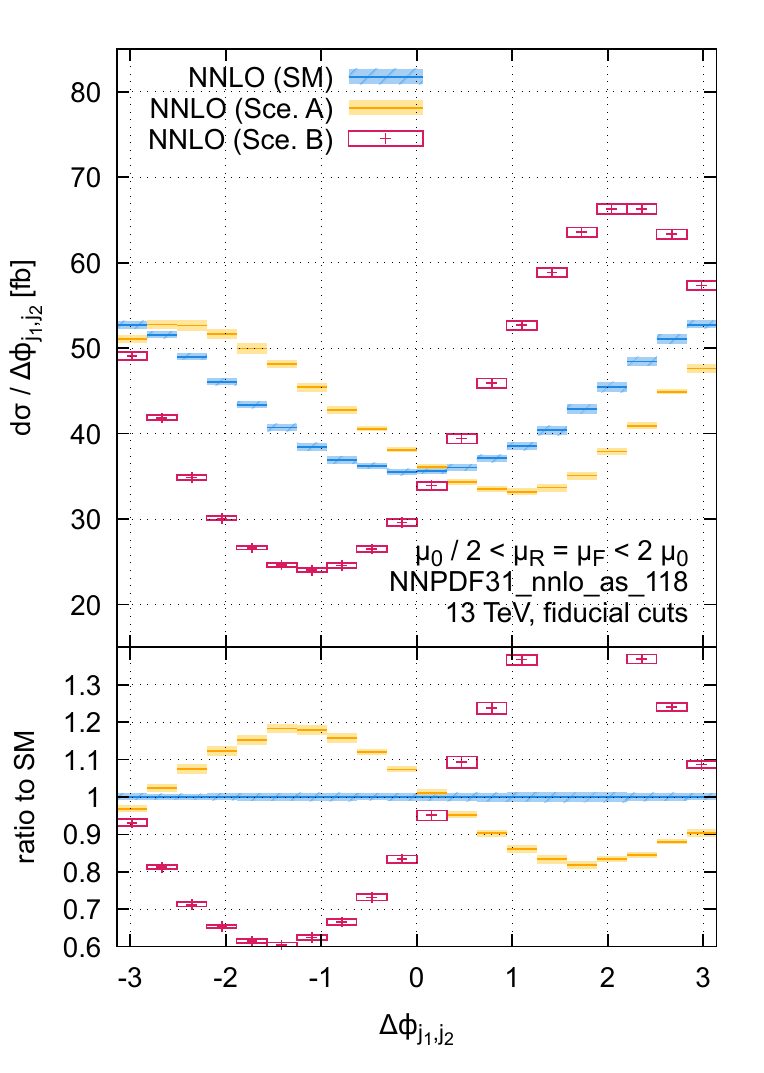}
	\caption{Same as Fig.~\ref{fig:histo:scea:pth} but for the
          azimuthal angle difference between the forward and backward
          VBF tag jets.  See text for details.}
	\label{fig:histo:dphi}
\end{figure*}

\begin{figure*}
	\centering
	\includegraphics[width=168pt]{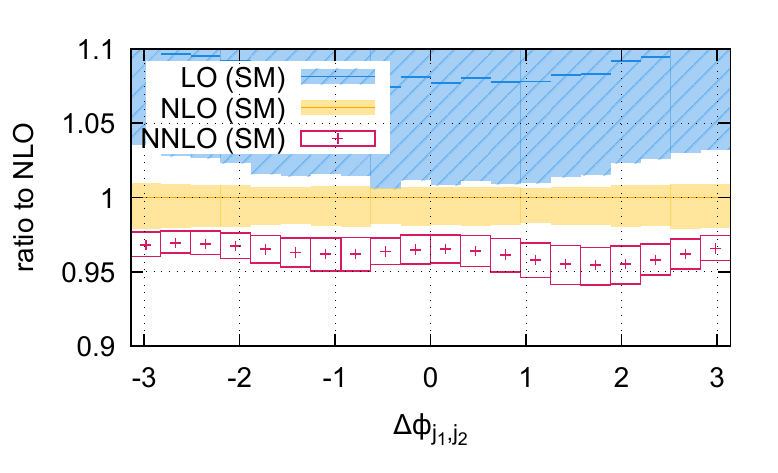}
	\includegraphics[width=168pt]{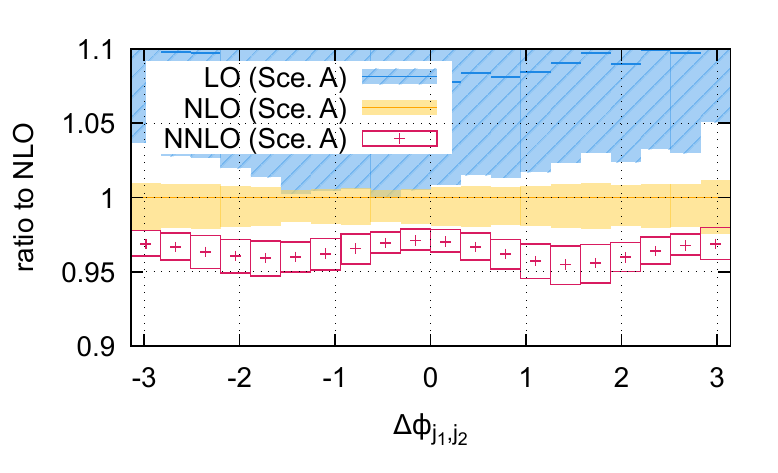}
	\includegraphics[width=168pt]{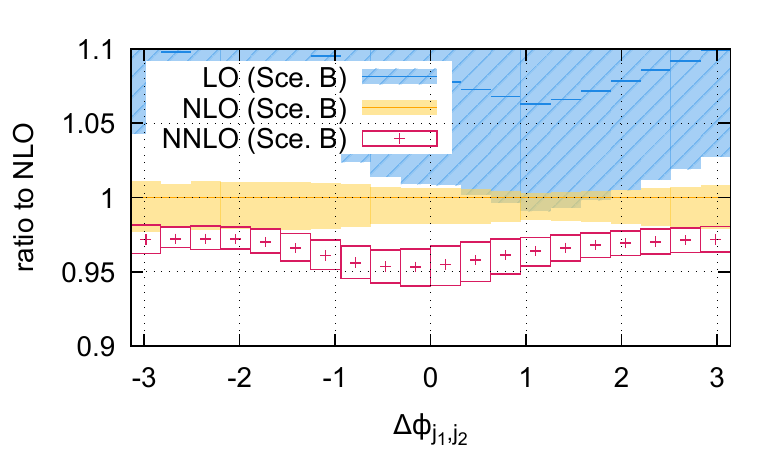}
	\caption{LO (hashed blue) and NNLO (red boxes) predictions for
          the azimuthal separation between the forward and backward
          VBF tag jets, divided by the corresponding NLO (solid
          yellow) result.  Left pane: Standard Model. Middle pane:
          Scenario A. Right pane: Scenario B. The envelope represents
          scale variation uncertainties.  See text for details.}
	\label{fig:histo:scenarios:dphi:kfact}
\end{figure*}

On the other hand, it is well-known that angular variables are
especially well-suited for discriminating between the SM and the
various EFT scenarios~\cite{Plehn:2001nj,Figy:2004pt,Hankele:2006ma}.
To this end, we study the azimuthal separation between the two hardest
jets $\Delta \phi_{j_1 j_2} = \phi_{j_1} - \phi_{j_2}$, where $j_1$ is
the forward jet and $j_2$ the backward jet, i.e. $y_{j_1} >0$ and
$y_{j_2} <0$.  We show this distribution in Fig.~\ref{fig:histo:dphi}.
It follows from this figure that, at LO, while the differences between
the SM and scenario A is just covered by scale variation
uncertainties, scenario B is clearly distinguishable from the SM as
well as from scenario A.  The situation improves at NLO.  Indeed, in
this case the scale uncertainty is significantly reduced and the three
scenarios become clearly distinguishable. The situation remains the
same at NNLO. We note that similar to the situation with the fiducial
cross section, there is no significant reduction in NNLO scale
variation uncertainties with respect to NLO.  It can also be seen in
Fig.~\ref{fig:histo:dphi} that the NLO shapes are quite stable under
radiative corrections.  This is a welcome feature, as it implies that
the perturbative expansion for this observable is under very good
control and NLO results are reliable. To quantify this statement, in
Fig.~\ref{fig:histo:scenarios:dphi:kfact} we plot the ratio of LO and
NNLO to NLO predictions, for both the Standard Model (left), scenario
A (middle) and scenario B (right). We see that in all the three cases
there is no overlap between the scale uncertainty bands, but the
corrections are rather flat and can be captured by a global
$K$-factor. Interestingly, both the very mild shape distortion and the
$K$-factor seem independent on the value of the anomalous couplings,
at least for the reference values chosen here. These results imply
that NLO predictions, possibly augmented by a global $K$-factor
rescaling, seem to provide a robust enough framework for performing
anomalous coupling studies with this observable.

\begin{figure*}
	\centering
	\includegraphics[width=168pt]{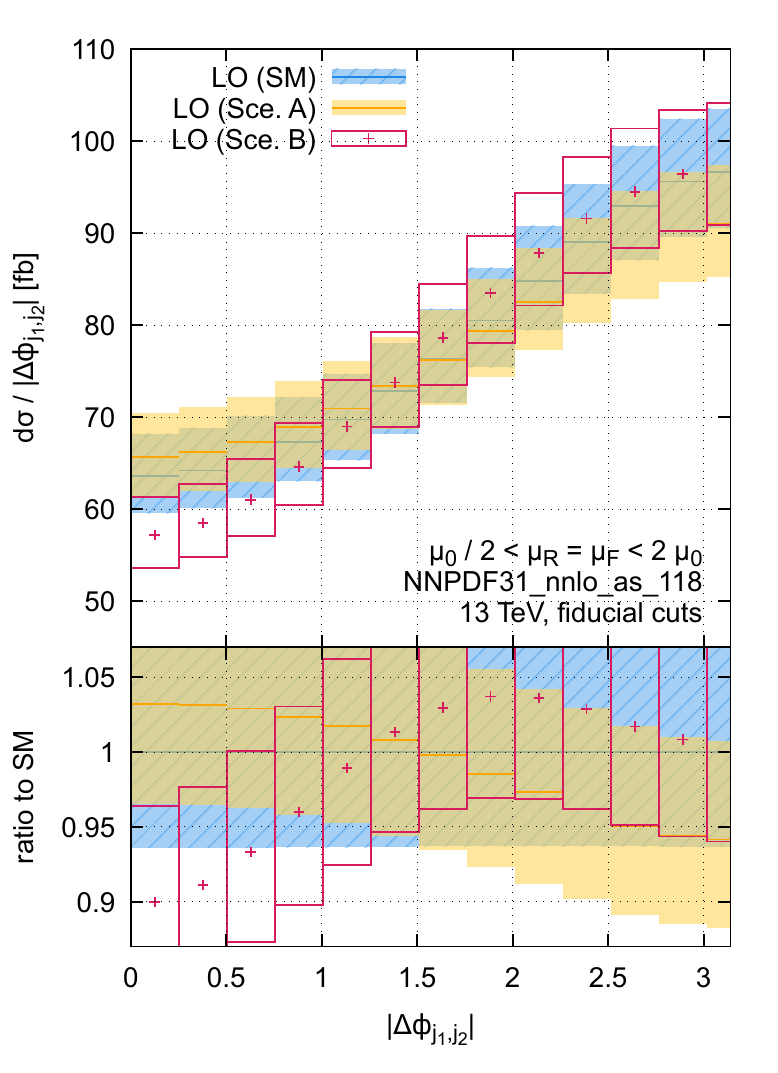}
	\includegraphics[width=168pt]{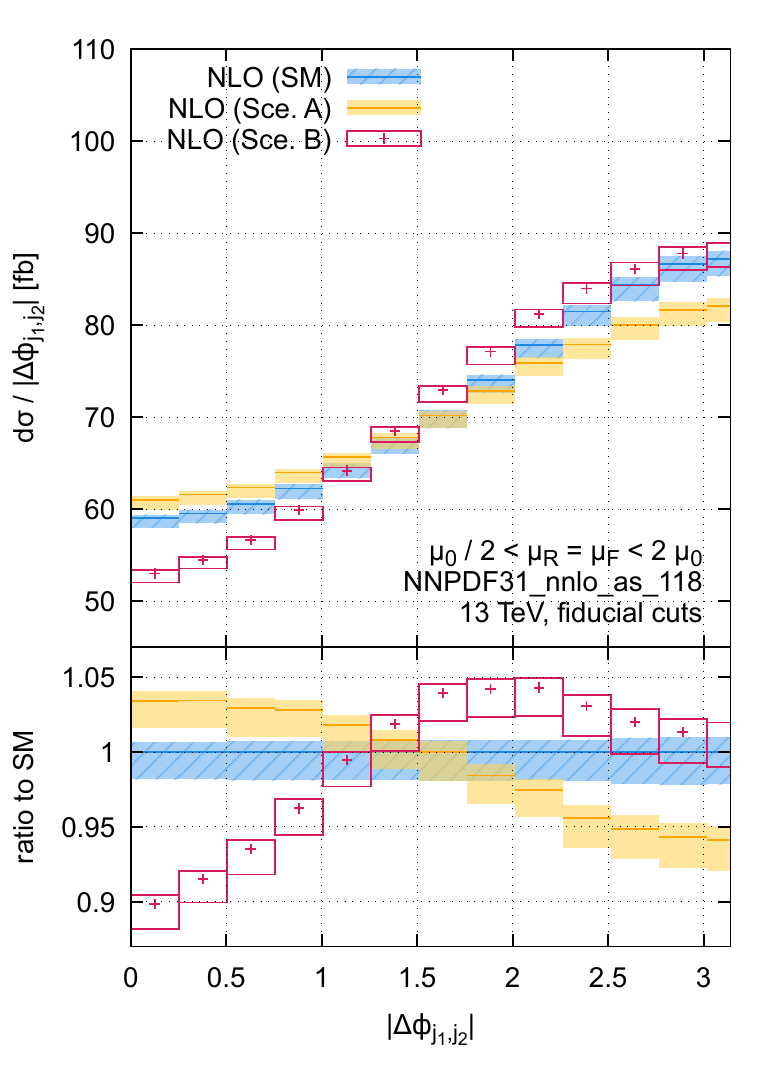}
	\includegraphics[width=168pt]{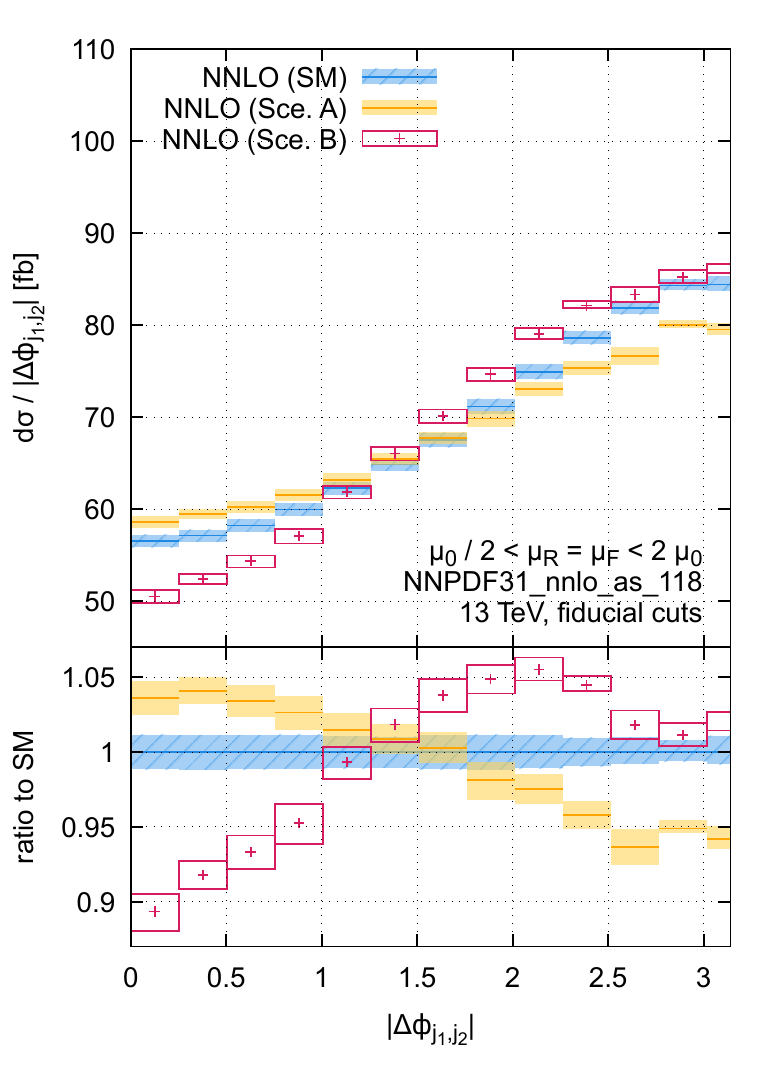}
	\caption{Same as Fig.~\ref{fig:histo:dphi} but for the
          absolute value of the azimuthal angle difference between the
          forward and backward VBF tag jets.  See text for details.}
	\label{fig:histo:abs_dphi}
\end{figure*}

\begin{figure*}
	\centering
	\includegraphics[width=168pt]{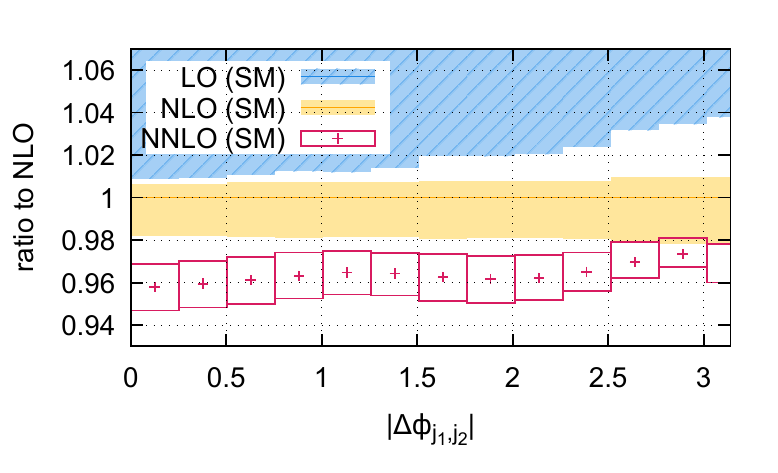}
	\includegraphics[width=168pt]{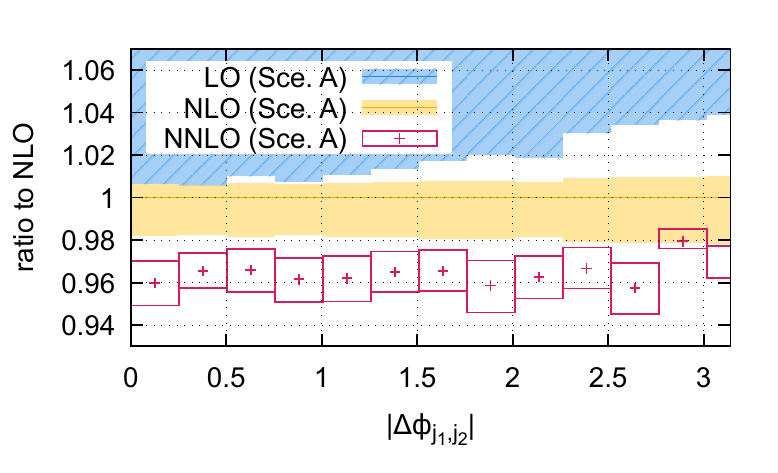}
	\includegraphics[width=168pt]{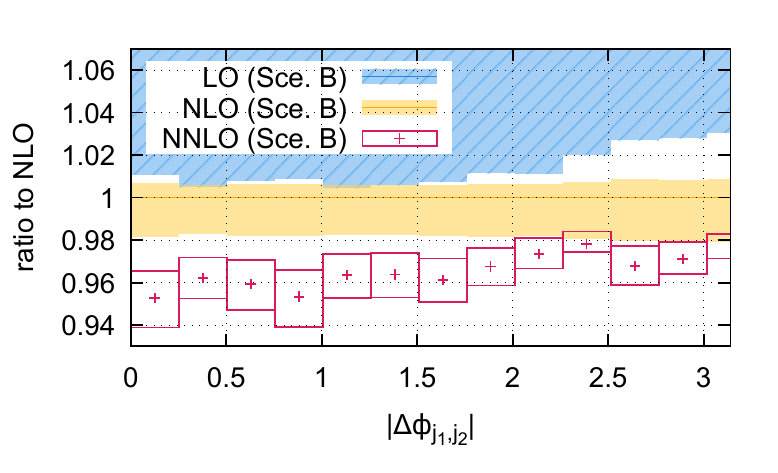}
	\caption{Same as Fig.~\ref{fig:histo:scenarios:dphi:kfact} but
          for the absolute azimuthal angle separation between the
          forward and backward VBF tag jets.  See text for details.}
	\label{fig:histo:abs_dphi:kfact}
\end{figure*}

Differences between scenario A, scenario B and the SM in the
distributions shown in Fig.~\ref{fig:histo:dphi} are primarily of an
anti-symmetric nature since they are dominated by the interference of
$CP$-odd and $CP$-even couplings, which are absent in
azimuthally-averaged observables.  Hence, in the presence of
non-vanishing $CP$-odd EFT operators, this observable may not be
optimal to study effects of $CP$-even couplings.  To this end, it is
useful to consider the absolute value of the azimuthal separation, in
which the interference between $CP$-odd and $CP$-even operators drops
out~\cite{Plehn:2001nj,Figy:2004pt,Hankele:2006ma}.  We show this
distribution in Fig.~\ref{fig:histo:abs_dphi}.

As expected, the deviations of scenarios A and B from the SM for
$|\Delta \phi_{j_1 j_2}|$ in Fig.~\ref{fig:histo:abs_dphi} are much
smaller than those observed in Fig.~\ref{fig:histo:dphi}.  In fact, at
LO, the differences between the various scenarios are entirely swamped
by scale variation uncertainties.  However, despite being less
sensitive to non-SM couplings, already at NLO scale uncertainties are
sufficiently reduced and the three scenarios become distinguishable.
Different shapes of scenarios A and B, due to different $CP$-even
contributions, can now be clearly observed.  Ratios in
Fig.~\ref{fig:histo:abs_dphi:kfact} imply that these shapes are also
relatively stable under radiative corrections and might, therefore, be
captured by global $K$-factors (as was the case for the $\Delta
\phi_{j_1 j_2}$ observable).

\section{Conclusions}
\label{sec:concl}

We have studied the combined impact of anomalous Higgs interactions
and NNLO QCD corrections on Higgs production in weak boson fusion. For
simplicity, we have only considered modifications to the $HVV$ vertex
and have performed phenomenological analyses assuming that the
modification of the $HWW$ and $HZZ$ couplings are correlated.  We have
found that the patterns of radiative corrections for the WBF cross
section in a typical experimental fiducial region are similar in the
Standard Model and when anomalous couplings are present. Namely, in
all cases scale variation uncertainties underestimate the size of the
correction at the next perturbative order, but predictions seem to be
reasonably stable when moving from NLO to NNLO. Keeping in mind that
for inclusive SM cross sections the NNLO scale uncertainty band does
overlap with N$^3$LO QCD predictions, we have investigated the effect
of a reduced theoretical uncertainty on extractions of anomalous
couplings. We have studied the constraining power of the fiducial
cross section, and found that NLO and NNLO results lead to a similar
discriminating power, which is significantly better than the LO one.
The relative stability of this picture, when going from NLO to NNLO,
gives us confidence that for the analysis of the anomalous couplings
in Higgs production through WBF, QCD radiative corrections are
sufficiently well understood.

We have also investigated the constraining power of differential
distributions, focusing on scenarios where different choices of
anomalous couplings led to the same fiducial cross section.  It is
known that angular distributions have a strong constraining power.
However, including NLO QCD corrections is mandatory for differences in
shapes induced by small anomalous couplings not to be swamped by
theoretical uncertainties.  We have shown that theoretical predictions
for differential distributions are reasonably stable when moving from
NLO to NNLO. In particular, we have found that shape distortions are
quite small, and the bulk of the effect of NNLO QCD corrections seems
to be captured by a global $K$-factor.  We have found that this holds
irrespective of the presence or absence of anomalous couplings, at
least for the scenarios studied in this paper.

There are several ways in which the analysis presented in this paper
can be improved.  First, one could study more general scenarios where
modifications of the $HZZ$ and $HWW$ couplings are different. Also,
one may study the impact of dimension-eight operators in $HVV$
couplings on Higgs signal in WBF.  Implementing both of these
extensions in our framework is conceptually straightforward. A more
involved improvement would be to consider a wider class of higher
dimensional operators, especially those that are affected by
higher-order QCD corrections.  Also, one may consider more realistic
setups, where e.g. the decay of the Higgs boson is also taken into
account. We look forward to investigating these and other issues
relevant for EFT studies in weak boson fusion in the future.


{\bf Acknowledgments:} We would like to thank Michael Trott for
helpful advice regarding the results of Ref.~\cite{Helset:2020yio}.
This research is partially supported by the Deutsche
Forschungsgemeinschaft (DFG, German Research Foundation) under grant
396021762 - TRR 257. The research of F.C. was partially supported by
the ERC Starting Grant 804394 {\sc hipQCD} and by the UK Science and
Technology Facilities Council (STFC) under grant ST/T000864/1. The
research of K.A. is supported by the United States Department of
Energy under Grant Contract DE-SC0012704.


\bibliography{anom}{}

\end{document}